# Multiple quantum NMR of spin-carrying molecules in nanopores: high order corrections to the two-spin/two-quantum Hamiltonian

**Sergei I. Doronin[a], Anna V. Fedorova[a], Edward B. Fel'dman[a], Alexander I. Zenchuk[a]**
DOI: 10.1039/b000000x

5   This paper is devoted to the multiple-quantum (MQ) NMR spectroscopy in nanopores filled by a gas of spin-carrying molecules ($s=1/2$) in a strong external magnetic field. It turned out that the high symmetry of the spin system in nanopores yields a possibility to overcome the problem of the exponential growth of the Hilbert space dimension with an increase in the number of spins and to investigate MQ NMR dynamics in systems
10  consisting of several hundred spins. We investigate the dependence of the MQ coherence intensities on their order (the profile of the MQ coherence intensities) for a spin system governed by the standard MQ NMR Hamiltonian (the nonsecular two-quantum/two-spin Hamiltonian) together with the second order correction of the average Hamiltonian theory. It is shown that the profile depends on the value of this correction and varies
15  from the exponential to the logarithmic one.

## 1 Introduction

Multiple-quantum NMR spectroscopy in solids [1] is widely used for the investigation of spin cluster growth in the process of irradiation of the spin system on the preparation period of
20 a MQ NMR experiment [1]. In particular, MQ NMR allows one to count the number of spins in clusters [2], to measure the decoherence rate for highly correlated spin states [3,4], to study the scaling of the decoherence rate with the number of correlated spins [3].

25 It is well known that the many-spin problems of MQ NMR dynamics represent the theoretical basis of MQ NMR spectroscopy. In particular, comparing the profiles of MQ NMR coherence intensities (i.e. the dependence of the MQ coherence intensities on their orders) at different times of the
30 irradiation of the system during the preparation period of the MQ NMR experiment with the theoretical curves one can obtain information about the growth of dipolar spin cluster size in the spin evolution process [1]. However, possibilities of the theoretical methods are restricted. The basic obstacle is
35 the exponential growth of the Hilbert space dimension with an increase in the number of spins, which may not be overcome in general. Nevertheless there are significant achievements in this direction.

The current state of affairs with the theoretical and
40 numerical methods of MQ NMR in solids is the following. The phenomenological approach [1], which is widely used for an interpretation of MQ NMR experiments, yields the Gaussian dependence of the MQ NMR coherence intensities on their orders (the Gaussian profile of MQ coherence
45 intensities). However, it is worth to note that this approach reduces the MQ problem to the combinatorial one and does not take into account the quantum-mecanical essence of the problem. Thus it becomes important to obtain profiles of MQ NMR coherence intensities using analytical and/or numerical
50 approaches instead of phenomenological one. In general, this problem is very complicated, so that, until recently, there was no exactly solvable model allowing one to make conclusions about profiles of MQ NMR coherence intensities. A simple example of the exactly solvable model is the one-dimensional
55 spin chain with the double quantum Hamiltonian [1] in the approximation of nearest neighbor interactions. It is shown that, starting with a thermodynamic equilibrium state, only zero and double quantum coherences are produced [5-7]. Next-nearest couplings and other distant interactions lead to
60 higher order coherences which can be observed in MQ NMR [1]. But these interactions are beyond the exactly solvable models [5-7]. As a result, the exactly solvable models do not allow us to investigate the profiles of MQ NMR coherence intensities. This means that one should concentrate on
65 numerical simulations in order to take into account distant spin-spin couplings. However the different numerical approaches to the problems of MQ NMR dynamics [8,9] allow us to simulate systems consisting of no more than twenty spins which is again insufficient in order to find the profiles
70 of MQ NMR coherence intensities.

Recently nanosize spin systems had a new impact on investigations of MQ NMR dynamics [10]. A closed nanopore filled by spin carrying atoms (molecules) is an example of such a system. It is shown that the MQ NMR dynamics of a
75 gas of spin carrying atoms (molecules) in a nanopore avoids the problem of the exponential growth of the Hilbert space dimension with an increase in the number of spins [10]. This happens due to the high symmetry of the MQ NMR Hamiltonian in nanopores which is associated with two
80 reasons.

The first reason is determined by a possibility of describing the dipole-dipole interactions (DDIs) in nanopores with only one coupling constant. The point is that the molecular diffusion does not completely average the DDIs of spin
85 carrying atoms (molecules) of a gas in the nanopores in a strong external magnetic field [11,12] only if the nanopores are nonspherical. Since the characteristic time scale of the molecular diffusion $t_{diff}$ is much less than the NMR time scale $t_d$ determined by the DDIs ($t_{diff}/t_d \approx 10^{-7}$) [12] one can
90 suppose that the residual averaged DDIs are determined by only one coupling constant, which is the same for all pairs of interacting spins [11,12]. We emphasize that the accuracy of this approximation is very high and any correction is senseless. As a result, the multiple quantum Hamiltonian commutes with the operator of the square of the total spin angular momentum and this integral of motion leads to the block structure of the Hamiltonian.

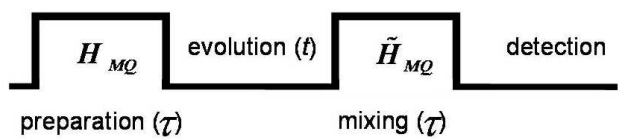

**Fig. 1** The basic scheme of the MQ NMR experiment when MQ dynamics is described by the corrected Hamiltonian.

The second reason is the permutative symmetry of the spin system in nanopores which yields further simplifications. A way of using this high symmetry in the problem of MQ NMR dynamics was developed in our recent article [10]. As a result, MQ NMR dynamics has been analyzed in systems consisting of hundreds of spins so that it became possible to find the profiles of MQ NMR coherence intensities.

The first experimental data on MQ NMR spectroscopy in solids [1] were in a good agreement with the conclusion of the phenomenological theory about the Gaussian profile of MQ NMR coherence intensities. However it was shown later [13] after processing a large amount of MQ NMR data that the profile is exponential. Ernst with coworkers [14] have found that the accuracy of MQ NMR experiment is insufficient to make a choice between Gaussian and exponential profiles. The investigation of MQ NMR dynamics of a gas of spin carrying atoms (molecules) in nanopores yields an unambiguous choice. The profile of MQ NMR coherence intensities turned out to be an exponential one [10]. It is found that the intensities of MQ NMR coherences of orders $4m+2$ and $4n$ ($m$, $n$ are integer) are described by different exponential curves in this model.

It is necessary to emphasize that all developed methods of MQ NMR dynamics use only the non-secular two-quantum/two-spin Hamiltonian in order to describe the behavior of spin systems in MQ NMR experiments. However higher order corrections of average Hamiltonian theory [15] can also be important for MQ dynamics. We demonstrate how the first nonzero correction affects on MQ NMR dynamics and study peculiarities of the profiles of MQ NMR coherence intensities associated with this correction.

We also remark that MQ NMR dynamics can be applied to an analysis of MQ NMR spectra [1]. Since the coupling constant averaged by the molecular diffusion, $D$, depends on the volume and shape of the nanopore and its orientation relatively to the external magnetic field [11,12], one can obtain this information from a comparison of the calculated MQ NMR spectra with experimental ones. It is also possible to obtain information about the number of spin-carrying atoms (molecules) in nanopores studying the profile of MQ NMR coherences. This is an advantage of MQ NMR in comparison with possibilities of standard NMR for investigations of nanopore compounds [11]. In addtition, MQ NMR dynamics of fluctuating nanocontainers allows us to obtain information about the correlation times of these fluctuations.

The paper is organized as follows. The theory of MQ NMR dynamics of spin carrying ($s=1/2$) atoms (molecules) in nanopores is developed in Sec. II. In particular, the second order correction of average Hamiltonian theory [15] to the two-spin/two-quantum Hamiltonian [1] is obtained. The block structure of the MQ NMR Hamiltonian with the obtained correction is described in Sec. III. The numerical profiles of the intensities of MQ NMR coherences are given in Sec. IV. An effect of corrections to the MQ NMR Hamiltonian on the profile of MQ NMR coherence intensities is discussed therein. We briefly summarize our results in the concluding Sec. V.

## 2 MQ NMR dynamics in systems of equivalent spins

### 2.1 Multiple quantum Hamiltonian with the second order correction

Let us consider a nanopore which is filled with a gas of spin-carrying ($s$=1/2) atoms (molecules). The DDI of spins in a strong external magnetic field is governed by the Hamiltonian

$$H_{dz} = \sum_{j<k} D_{jk}[2I_{jz}I_{kz} - \frac{1}{2}(I_j^+ I_k^- + I_j^- I_k^+)], \qquad (1)$$

where $D_{jk} = \gamma^2 \hbar(1-3\cos^2\theta_{jk})/(2r_{jk}^3)$ is the DDI coupling constant between spins $j$ and $k$, $\gamma$ is the gyromagnetic ratio, $r_{jk}$ is the distance between spins $j$ and $k$, and $\theta_{jk}$ is the angle between the intermolecular vector $\vec{r}_{jk}$ and the external magnetic field $\vec{H}$ which is parallel to $z$ axis [16]. The operator $I_{j\alpha}$ ($\alpha=x, y, z$) is the projection of the spin angular momentum operator on the axis $\alpha$; $I_j^+$ and $I_j^-$ are the raising and lowering operators of spin $j$.

A standard MQ NMR experiment can be divided into four distinct periods of time (Fig. 1): preparation ($\tau$), evolution ($t$), mixing ($\tau$), and detection. The preparation period is the heart of the MQ NMR experiment. MQ coherences are created by a periodic multipulse sequence (with period $\tau_c$) consisting of eight-pulse cycles which irradiates the spin system during the preparation period [1]. The eight-pulse cycles consist of resonance $(\pi/2)_{\pm x}$ pulses with alternating delays $\delta$ and $2\delta$, so that $\tau_c = 12\delta$. In the rotation reference frame [16] the DDI Hamiltonian (1) averaged by the multipulse sequence, $H_{MQ}$, can be presented according to the average Hamiltonian theory [15] in the following form

$$H_{MQ} = H_0 + H_2, \qquad (2)$$

where $H_0$ is the averaged nonsecular two-spin/two-quantum Hamiltonian

$$H_0 = H^{(2)} + H^{(-2)}, \qquad (3)$$

$$H^{(\pm 2)} = -\frac{1}{2}\sum_{j<k} D_{jk} I_j^\pm I_k^\pm, \qquad (4)$$

and $H_2$ is the first non-zero correction to the average Hamiltonian $H_0$ [15,17]:

$$H_2 = \frac{\tau_c^2}{36}\left[\left[H_{dy}, H_{dx}\right], H_{dx} - 3H_{dy}\right] \qquad (5)$$

where $H_{dx}$ ($H_{dy}$) can be obtained from Eq. (1) by the replacement $z \leftrightarrow x$ ($z \leftrightarrow y$).

The derived expressions (3), (4) for $H_0$ and (5) for $H_2$ are valid for any spin system. However, the Hamiltonian of spin

**2** | *Journal Name*, [year], **[vol]**, 00–00                                This journal is © The Royal Society of Chemistry [year]

system in nanopores allows significant simplification. This becomes possible due to the remarkable fact that the characteristic time of the molecular spin diffusion in a nanopore $t_{diff}$ is about $10^{-11}$ s, which is much less than the characteristic DDI time $t_d \approx 10^{-4}$ s [12]. Thus one can suppose that the averaged DDI coupling constant $D$ is the same for all spin pairs [11,12]. Since the MQ NMR Hamiltonian is described by only one coupling constant $D$ one can rewrite Eqs. (3) and (4) as follows [10]:

$$H_0 = -\frac{D}{4}\{(I^+)^2 + (I^-)^2\}, \quad (6)$$

where $I^\pm = \sum_{j=1}^{N} I_j^\pm$ and $N$ is the number of spins in the nanopore. Analogously the correction $H_2$ of Eq. (5) can be written as follows

$$H_2 = D^2 \tau_c^2 \{\frac{3}{4}I_z + \frac{3}{8}(I^-)^2 + \frac{3}{8}(I^+)^2 + \frac{9}{4}I_z^2 - \frac{3}{4}(I^-)^2 I_z$$
$$-\frac{3}{2}I^+I^-I_z + \frac{3}{4}(I^+)^2 I_z + \frac{3}{2}I_z^3 - \frac{3}{16}(I^-)^4 + \frac{3}{8}(I^-)^2 I_z^2 \quad (7)$$
$$-\frac{3}{2}I^+I^-I_z^2 + \frac{3}{8}(I^+)^2(I^-)^2 - \frac{3}{16}(I^+)^4 + \frac{3}{8}(I^+)^2 I_z^2\}.$$

At first glance, the smallness of the correction $H_2$ given by Eq. (7) is defined by the parameter $\varepsilon = D\tau_c$. However, it is not difficult to show that the smallness of this correction is defined by the different parameter. In fact, it is well known [15] that the local dipolar frequency at every spin should be smaller than the inverse period (which is proportional to $\tau_c^{-1}$) of the multipulse sequence irradiating the spin system during the preparation period. This requirement is necessary in order for the application of average Hamiltonian theory to be valid. The local dipolar frequency is proportional to the square root of the second moment of the NMR absorption line, $\sqrt{M_2}$. In turn, this second moment, $M_2$, is proportional to the square of the dipolar coupling constant $D$, and to the number of nearest neighbors associated with the given spin. Since all spins are nearest neighbors in our model one can obtain that the smallness of the correction $H_2$ is defined by the parameter $\varepsilon_N$:

$$\varepsilon_N = \varepsilon\sqrt{N} = D\tau_c\sqrt{N}. \quad (8)$$

To estimate the value of $\varepsilon_N$ we refer to the experimental data of Ref. 11 in hydrogenated amorphous silicon (a-Si:H) with nanovoids that contain high-pressure $H_2$ gas. In accordance with these data, the DDI coupling constant averaged by the molecular diffusion is $D \approx 2\pi \cdot 300$ Hz. Taking the time spacing between two pulses of the irradiating multipulse sequence $\tau_c \approx 20$ μs one can find that $\varepsilon_N \approx 0.53$ if the number of spins $N = 200$. This estimation shows that the parameter $\varepsilon_N$ is not small in general. Consequently, the correction $H_2$ given by Eq. (7) should be taken into account in analysis of the experimental data of MQ NMR.

## 2.2 Intensities of MQ NMR coherences

In order to find expressions for the intensities, first of all we have to find the longitudinal polarization $<I_z>(\tau,t)$ after the mixing period of the MQ NMR experiment (Fig.1). Starting with the thermodynamic equilibrium state one can find

$$<I_z>(\tau,t) = \text{Tr}\{V^+(\tau)e^{-i\Delta t I_z}U(\tau)I_z U^+(\tau)e^{i\Delta t I_z}V(\tau)I_z\} =$$
$$\text{Tr}\{e^{-i\Delta t I_z}U(\tau)I_z U^+(\tau)e^{i\Delta t I_z}V(\tau)I_z V^+(\tau)\} \quad , \quad (9)$$

where we introduce evolution operators $U(\tau)$ for the preparation period and $V^+(\tau)$ for the mixing period as follows:

$$U(\tau) = \exp(-iH_{MQ}\tau), \quad V^+(\tau) = \exp(-i\tilde{H}_{MQ}\tau). \quad (10)$$

Here the Hamiltonian $H_{MQ}$ is defined by Eq. (2) and the Hamiltonian $\tilde{H}_{MQ}$ can be written in the following form

$$\tilde{H}_{MQ} = e^{i\frac{\pi}{2}I_z}H_{MQ}e^{-i\frac{\pi}{2}I_z} = -H_0 + \tilde{H}_2 \quad (11)$$

where the correction $\tilde{H}_2$ is

$$\tilde{H}_2 = e^{i\frac{\pi}{2}I_z}H_2 e^{-i\frac{\pi}{2}I_z}. \quad (12)$$

Note that the transformation from $H_{MQ}$ to $\tilde{H}_{MQ}$, i.e. Eq. (11), is equivalent to the replacement $x \leftrightarrow y$. As a result, $H_0$ just changes sign, while transformation of $H_2$ is more complicated, see Eq. (12). The offset $\Delta$ on the evolution period in Eq. (9) encodes MQ NMR coherences of different orders.

Next, it is useful to introduce the density matrices $\rho(\tau)$ and $\tilde{\rho}(\tau)$ in the preparation and mixing periods

$$\rho(\tau) = U(\tau)I_z U^+(\tau), \quad \tilde{\rho}(\tau) = V(\tau)I_z V^+(\tau), \quad (13)$$

which may be expanded in the series as follows:

$$\rho(\tau) = \sum_n \rho_n(\tau), \quad \tilde{\rho}(\tau) = \sum_n \tilde{\rho}_n(\tau), \quad (14)$$

where $\rho_n(\tau)$, $\tilde{\rho}_n(\tau)$ are the contributions to $\rho(\tau)$ and $\tilde{\rho}(\tau)$ from the coherences of $n$th order. Using Eqs. (9), (13), (14) and taking into account that [6]

$$e^{-i\Delta t I_z}\rho_n(\tau)e^{i\Delta t I_z} = e^{-in\Delta t}\rho_n(\tau),$$
$$e^{-i\Delta t I_z}\tilde{\rho}_n(\tau)e^{i\Delta t I_z} = e^{-in\Delta t}\tilde{\rho}_n(\tau), \quad (15)$$

one can obtain

$$<I_z>(\tau,t) = \sum_n e^{-in\Delta t}\text{Tr}\{\rho_n(\tau)\tilde{\rho}_m(\tau)\}. \quad (16)$$

Finally, since the trace operation is invariant with respect to rotations about the $z$ axis one can find that

$$<I_z>(\tau,t) = \sum_n e^{-in\Delta t}\text{Tr}\{\rho_n(\tau)\tilde{\rho}_{-n}(\tau)\} \quad (17)$$

and, as a consequence, the intensity $J_n(\tau)$ of the MQ NMR coherence of order $n$ is determined by

$$J_n(\tau) = \text{Tr}\{\rho_n(\tau)\tilde{\rho}_{-n}(\tau)\} \quad (18)$$

Now let us show that the intensities $J_n(\tau)$ are real. Using Eqs. (10) and (13) one can find that

$$\tilde{\rho}(t) = e^{-i\frac{\pi}{2}I_z}\rho^*(t)e^{i\frac{\pi}{2}I_z} \quad (19)$$

and, consequently,

$$\tilde{\rho}_n(t) = e^{-i\frac{\pi}{2}n}\rho^*_n(t). \quad (20)$$

Next, using Eq.(9), making the unitary transformation of the expression $e^{-i\Delta t I_z}U(\tau)I_z U^+(\tau)e^{i\Delta t I_z}V(\tau)I_z V^+(\tau)$ with the operator $\exp(-i\pi I_x)$ and using Eqs.(13)-(15) one can find that

$$<I_z>(\tau,t)=\sum_n e^{in\Delta t}J_n(\tau).\quad(21)$$

Since the RHS of Eqs.(16) and (21) must coinside, one obtains

$$\sum_n \sin(\Delta nt)J_n(\tau)=0.\quad(22)$$

Eq. (22) leads us to the obvious relation

$$J_n(\tau)=J_{-n}(\tau),\quad(23)$$

which allows us to conclude that the intensities $J_n(\tau)$ are real. However the intensities $J_n(\tau)$ can be negative [18] and their sum is not time invariant as in the standard MQ NMR dynamics [19].

## 3 The block structure of the MQ NMR Hamiltonian for a spin system in a nanopore

In this section we discuss some symmetries of the Hamiltonian $H_{MQ}$, which provide significant simplifications of numerical simulations.

In general, the dimension of the MQ NMR Hamiltonian, $H_{MQ}$, of Eq. (2) is $2^N$ and it grows exponentially with an increase in the number of spins $N$. Usually eigenstates of $I_z$ (the projection of the total spin angular momentum on the external magnetic field) are used as a basis in order to describe MQ NMR dynamics [1] (the multiplicative basis). Then the blocks with binomial dimensions appear in calculations of intensities of MQ NMR coherences. At the same time, the Hamiltonian $H_{MQ}$ of Eq. (2) commutes with the total spin angular momentum $\hat{I}^2$ and it is suitable to investigate MQ NMR dynamics in a nanopore using the basis of common eigenstates of $\hat{I}^2$ and $I_z$ [10]. In order to obtain the matrix representation of $H_{MQ}$ one should take into account that the maximal total spin number is $I=N/2$, and its different values are $S=N/2, N/2-1, \ldots, N/2-[N/2]$, where $[a]$ is an integer part of $a$ [20]. It is also important that nonzero elements of the matrix representations of the raising $I^+$ and lowering $I^-$ operators corresponding to the total spin number $S$ are [20]

$$<M|S^+|M-1>=<M-1|S^-|M>=\sqrt{(S+M)(S-M+1)},\quad(24)$$

where $M=-S+1, -S+2, \ldots, S-1, S$. Since the Hamiltonian $H_0$ and the correction $H_2$ consist of the operators $I^+$, $I^-$, and $I_z$, Eq. (24) is sufficient in order to obtain the matrix representation of the MQ Hamiltonian $H_{MQ}$. The number of the energy levels, $n_N(S)$, corresopnding to the total momentum $S$ in an $N$-spin system is [20]

$$n_N(S)=\frac{N!(2S+1)}{(\frac{N}{2}+S+1)!(\frac{N}{2}-S)!},\quad 0\le S\le \frac{N}{2}\quad(25)$$

which means that the block of the MQ Hamiltonian corresponding to the total spin number $S$ is degenerate and has the dimension $2S+1$. One can verify that [10]

$$\sum_S n_N(S)(2S+1)=2^N.\quad(26)$$

Thus we have built the total basis consisting of the $2^N$ orthonormal vectors which allows us to represent the MQ Hamiltonian, $H_{MQ}$, as a set of blocks corresponding to different total spin numbers.

It is important that the initial density matrix is a thermodynamic equilibrium one and is equal to the matrix representation of the operator $I_z$. As a result, the problem is reduced to a set of the analogous tasks for each block $H_{MQ}^S$ ($S=N/2$, $S=N/2-1$, …, $N/2-[N/2]$) and the contribution $J_{k,S}(\tau)$ to the intensity of the $k$-th order MQ NMR coherence is determined as follows

$$J_{k,S}(\tau)=\frac{\text{Tr}\{\rho_k^S(\tau)\tilde{\rho}_{-k}^S(\tau)\}}{\text{Tr}\{I_z^2\}},\quad(27)$$

$$\rho_k=diag(\rho_k^{N/2},\rho_k^{N/2-1},\ldots,\rho_k^{N/2-[N/2]}),$$

$$\tilde{\rho}_k=diag(\tilde{\rho}_k^{N/2},\tilde{\rho}_k^{N/2-1},\ldots,\tilde{\rho}_k^{N/2-[N/2]}),$$

where $\rho_k^S(\tau)$ is the part of $\rho^S(\tau)$ which is responsible for the MQ NMR coherence of order $k$ and $\tilde{\rho}_{-k}^S$ is the corresponding part of the density matrix $\tilde{\rho}^S(\tau)$. Taking into account the multiplicity of the intensities $J_{k,S}(\tau)$ according to Eq. (25) one can obtain the observable intensities of MQ NMR coherences $J_k(\tau)$ ($-N\le k\le N$) in the following form

$$J_k(\tau)=\sum_S n_N(S)J_{k,S}(\tau).\quad(28)$$

It is also suitable to use one more symmetry of the MQ Hamiltonians of Eqs. (2), (11). The point is that the Hamiltonians $H_{MQ}$ and $\tilde{H}_{MQ}$ and all their blocks are invariant with respect to the rotation by an angle $\pi$ about the $z$ axis. It means that the operator $\exp(i\pi I_z)$ is an integral of motion for MQ NMR dynamics even when the corrections in Eqs. (2), (11) are taken into account. Thus $2^N\times 2^N$ Hamiltonian matrix is reduced to two $2^{N-1}\times 2^{N-1}$ matrices. Consequently, all blocks $H_{MQ}^S$ and $\tilde{H}_{MQ}^S$ ($S=N/2$, $N/2-1$, …, $N/2-[N/2]$) also split in two subblocks. It turns out that these submatrices yield the complex conjugate contribution into intensities of the MQ NMR coherences if only $N$ is odd [8]. For this reason, it is sufficient to perform calculations with only one submatrix and double the real parts of the resulting intensities. All numerical calculations in Sec.4 are performed for an odd numbers of spins.

The numerical algorithm for MQ NMR spin dynamics [7,8,10] was modified in the present work in order to take into account the corrections to the MQ NMR Hamiltonians for the preparation and mixing periods and the new expression of Eq. (18) for the intensities of MQ NMR coherences.

## 4 The numerical analysis of MQ NMR dynamics of a spin system with the corrected Hamiltonian in a nanopore

Performing numerical calculations one has to take into account that the correction $H_2$ of Eq. (5) depends on the small parameter $\varepsilon_N$, which depends on both the time spacing $\tau_c$ and the number of spins $N$ according to Eq.(8). We see that in order to satisfy the condition $\varepsilon_N<1$ one should decrease the time spacing $\tau_c$ with an increase in the number of spins $N$, which was used in numerical simulations. It is obvious that profiles of MQ coherence intensities must depend

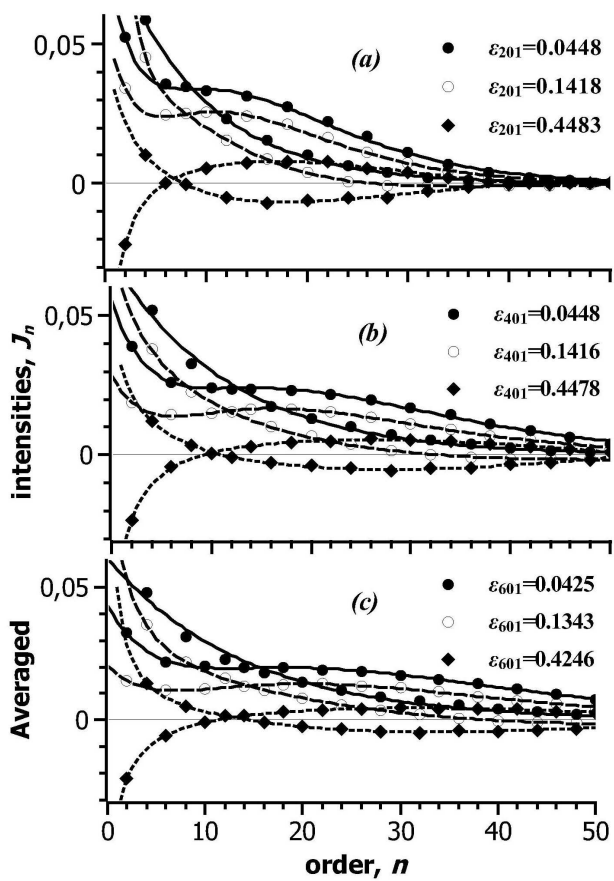

**Fig. 2** Profiles of the intensities of MQ coherences $\bar{J}_n$ averaged over the time interval $t_0 \le t \le t_0 + 2T$ with $t_0 = 31$ and $T = 7.255$ corresponding to the different values of $N$ and $\varepsilon_N$: (a) $N=201$, (b) $N=401$, (c) $N=601$

significantly on the value of the parameter $\varepsilon_N$, which is justified below.

Following the strategy of Ref. [10], we consider the averaged intensities $\bar{J}_k$,

$$\bar{J}_k = \frac{1}{2T}\int_{t_0}^{t_0+2T} J_k(t)dt, \quad t_0 = 31, \quad T = 4\pi/\sqrt{3} \approx 7.255.. \quad (29)$$

The dimensionless time $t = D\tau$ is used in Eq.(29) and below. Our choice of $t_0$ is motivated by the requirement that the quasistationary distribution of intensities is realized; $T$ must be long enough so that all intensities are quickly oscillating over the period $T$. We take the same values of $t_0$ and $T$ as in [10]. Similar to [10], all intensities are separated into two families

$$\begin{aligned}\Gamma_1 &= \{\bar{J}_{4k-2}, k=1,2,...\},\\ \Gamma_2 &= \{\bar{J}_{4k}, k=1,2,...\},\end{aligned} \quad (30)$$

while $\bar{J}_0$ may not be referred to any of these families. Each family is approximated by a smooth function. Profiles of MQ coherence intensities for $N = 201$, 401, and 601 spins with different values of $\varepsilon_N$ are represented in Fig. 2. This figure justifies the statement that the smallness of the correction $H_2$ is defined by the parameter $\varepsilon_N$ given by Eq.(8). In fact, if $\varepsilon_{201} \approx \varepsilon_{401} \approx \varepsilon_{601}$ then profiles corresponding to different $N$ have the same shape. Constructing the profiles, we consider only intensities $J_{2n}$ of orders $n=0,\pm 1,...,\pm 25$, since higher order intensities are negligeable. If $\varepsilon_N = 0$, then we have the intensity profiles obtained in [10]:

$$\bar{J}_k = \begin{cases} A_1(1-a_1|k|+a_2 k^2)\exp(-\alpha_1|k|), \\ \qquad k = \pm 2, \pm 6,... \\ A_2 \exp(-\alpha_2|k|), \quad k = \pm 4, \pm 8,... \end{cases} \quad (31)$$

The parameters $A_i$, $\alpha_i$, and $a_i$ ($i = 1,2$) have been found for $N = 201$, 401, and 601, see Table 1.

**Table 1** The parameters in Eq. (31) corresponding to different values of $N$ and $\varepsilon_N = 0$

| $N$ | $\bar{J}_0$ | $A_1$ | $A_2$ | $\alpha_1$ | $\alpha_2$ | $a_1$ | $a_2$ |
|---|---|---|---|---|---|---|---|
| 201 | 0.1973 | 0.0875 | 0.0912 | 0.1676 | 0.1140 | 0.1296 | 0.0235 |
| 401 | 0.1604 | 0.0560 | 0.0704 | 0.1154 | 0.0858 | 0.0866 | 0.0120 |
| 601 | 0.1414 | 0.0437 | 0.0608 | 0.0931 | 0.0733 | 0.0691 | 0.0082 |

If $\varepsilon_N$ is small, then the correction $H_2$ to the Hamiltonian is not significant and profiles of MQ NMR coherence intensities are only slightly deformed. Namely, they are described by Eq. (31) with the parameters represented in Table 2.1. These profiles are shown in Fig.2 by solid lines.

**Table 2.1** The parameters in Eq. (31) corresponding to different values of $N$; $\varepsilon_{201} = \varepsilon_{401} = 0.0448$, $\varepsilon_{601} = 0.0425$

| $N$ | $\bar{J}_0$ | $A_1$ | $A_2$ | $\alpha_1$ | $\alpha_2$ | $a_1$ | $a_2$ |
|---|---|---|---|---|---|---|---|
| 201 | 0.1970 | 0.0854 | 0.0916 | 0.1651 | 0.1148 | 0.1209 | 0.0227 |
| 401 | 0.1603 | 0.0551 | 0.0702 | 0.1125 | 0.0846 | 0.0799 | 0.0113 |
| 601 | 0.1414 | 0.0430 | 0.0605 | 0.0901 | 0.0717 | 0.0632 | 0.0076 |

**Table 2.2** The parameters $P_i$ ($i=1,2$) and $P$ corresponding to different values of $N$; $\varepsilon_{201} = \varepsilon_{401} = 0.0448$, $\varepsilon_{601} = 0.0425$

| $N$ | $P_1$ | $P_2$ | $P$ |
|---|---|---|---|
| 201 | 0.4887 | 0.3131 | 0.9988 |
| 401 | 0.4787 | 0.3427 | 0.9817 |
| 601 | 0.4524 | 0.3527 | 0.9465 |

However, the numerical results may not be approximated by Eqs. (31) if $\varepsilon_N$ is larger. In this case one has to look for more complicated curves. For instance, to approximate profiles associated with the middle value of the parameter $\varepsilon_N$ shown in Fig. 2 (dashed lines), Eq. (31) must be replaced with the following one:

$$\bar{J}_k = \begin{cases} A_1(1-a_{11}|k|+a_{12}k^2)\exp(-\alpha_1|k|), \\ \qquad k = \pm 2, \pm 6,... \\ A_2(1+a_{21}\sqrt{|k|}+a_{22}|k|+a_{23}|k|^{1/3})\exp(-\alpha_2|k|), \\ \qquad k = \pm 4, \pm 8,... \end{cases} \quad (32)$$

where the parameters are given in Table 3.1. Note that the intensities of MQ NMR coherences can be negative when $\varepsilon_N$ is not small (see Fig.2, dashed and dotted lines). Although this is a terminological problem only [18] an appearance of negative intensities means that the correction $H_2$ becomes quite important.

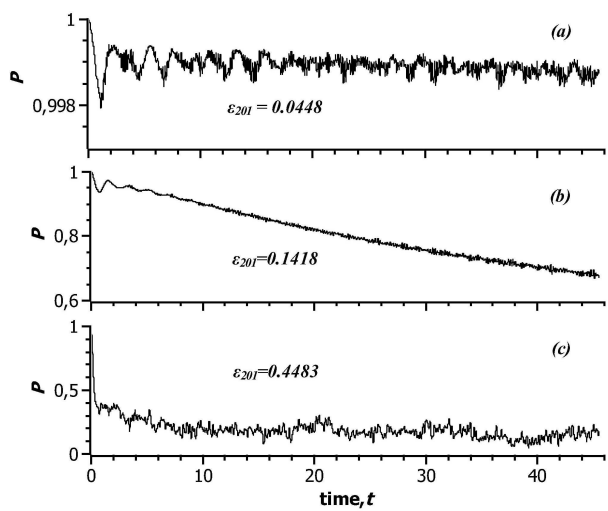

**Fig. 3** The evolution of the sum P= $\sum_{n=-50}^{50} \bar{J}_n$ for the system with $N$=201 and different values of $\varepsilon_{201}$

Regarding the biggest values of $\varepsilon_N$ in Fig.2 (dotted lines), the exponential function in Eq. (32) must be replaced with the logarithmic one. However, we do not represent analytical formulaes for the profiles in this case because such profiles mean that the effect of $H_2$ is too big and corrections of higher order of the average Hamiltonian theory [15,17] should be also involved.

Calculating the parameters given in Tables 2.1 and 3.1, we took into account the fact that $\sum_{n=-\infty}^{+\infty} J_n \neq 1$ (see Fig. 3). The point is that the conditions of a time reversal are not valid for the mixing period of the MQ NMR experiment if $\varepsilon_N \neq 0$. As a result, the intensities $J_n(\tau)$ of the MQ NMR conerences of Eq. (18) contain contributions from different density matrices and the sum of these intensities is not determined by the time invariant value

$$\text{Tr}\,\rho^2(\tau) = \sum_n \text{Tr}\{\rho_n \rho_{-n}\}, \quad (33)$$

unlike the case $\varepsilon_N = 0$ [19]. For this reason, we replace normalization $\sum_{n=-\infty}^{+\infty} J_n = 1$ (which has been used in ref.[10]) with the following one:

$$\sum_{n=1}^{25} 2\bar{J}_{2n} = P_1, \quad (34)$$

$$\sum_{n=1}^{23} 2\bar{J}_{2n+2} = P_2, \quad (35)$$

$$P = P_1 + P_2 + \bar{J}_0. \quad (36)$$

**Table 3.1** The parameters in Eq. (32) corresponding to different values of $N$ and $\varepsilon_N$

| $N$ | $\varepsilon_N$ | $\bar{J}_0$ | $A_1$ | $\alpha_1$ | $a_{11}$ | $a_{12}$ |
|---|---|---|---|---|---|---|
| 201 | 0.1418 | 0.1782 | 0.0611 | 0.1778 | 0.1657 | 0.0314 |
| 401 | 0.1416 | 0.1384 | 0.0285 | 0.1209 | 0.1104 | 0.0189 |
| 601 | 0.1343 | 0.1213 | 0.0203 | 0.0959 | 0.0828 | 0.0131 |

| $N$ | $A_2$ | $\alpha_2$ | $a_{21}$ | $a_{22}$ | $a_{23}$ |
|---|---|---|---|---|---|
| 201 | 1.5221 | 0.1331 | 1.3272 | -0.0560 | -2.1290 |
| 401 | 0.0936 | 0.0150 | -0.4136 | 0.0563 | -0.0025 |
| 601 | 0.1267 | 0.0655 | -0.5501 | 0.1417 | -0.0126 |

**Table 3.2** The parameters $P_i$ ($i$ = 1, 2) and $P$ corresponding to different values of $N$ and $\varepsilon_N$

| $N$ | $\varepsilon_N$ | $P_1$ | $P_2$ | $P$ |
|---|---|---|---|---|
| 201 | 0.1418 | 0.3455 | 0.1893 | 0.7130 |
| 401 | 0.1416 | 0.2982 | 0.1875 | 0.6240 |
| 601 | 0.1343 | 0.2800 | 0.2058 | 0.6071 |

Eqs. (34), (35) have been used to express $A_1$ and $A_2$ in terms of other parameters. Numerical values of $P_i$ ($i$ = 1, 2) and $P$ for different cases are given in Tables 2.2 and 3.2. Other parameters in Eq. (32), $a_{ij}$, $\alpha_i$, $i$ = 1,2, $j$ = 1,2,3, may be found using the least square method.

In conclusion of this section we remark that Fig.3 demonstrates that the sum of the MQ NMR intensities decreases more quickly when the parameter $\varepsilon_N$ is larger. A possible way to reduce this decrease is decreasing $\varepsilon_N$ (i.e. $\tau_c$) during the course of the MQ NMR experiment.

## Conclusions

We study MQ NMR dynamics of a spin system in a nanopore when the Hamiltonian, governing the spin dynamics, contains the correction term. This correction is determined by the second order correction of average Hamiltonian theory [15,17] and it crucially affects the profile of MQ NMR coherence intensities. We have found that the correction term conserves the symmetry of intensities of MQ NMR coherences with respect to a change in sign of the coherence order. However, the conservation law of the sum of MQ NMR coherences [19] is not valid and the intensities of MQ NMR coherences can be negative [18] for the sufficiently large parameter $\varepsilon_N < 1$. It is reasonable to assume that an emergence of negative intensities means that the time spacing $\tau_c$ should be decreased in MQ NMR experiments.

We emphasize that the small parameter $\varepsilon_N$ can be made independent on the number of spins in nanopores with the corresponding choice of the time spacing $\tau_c$. This fact can be used in order to find a set of parameters which are necessary for performing MQ NMR experiments. We also assume that the different curves for profiles of MQ NMR coherence intensities of orders $4n + 2$ and $4m$ ($n$, $m$ are integer) may serve as an indicator of an existence of only one dipolar coupling constant describing MQ dynamics.

Finally, we notice that the profile of MQ NMR coherence intensities is exponential for very small parameters $\varepsilon_N$. Otherwise the profile is described by more complex functions. It is worth to note that the exact solutions for spin dynamics in nanopores [12,21] are very promising for an interpretation of MQ NMR experiments.

## Acknowledgments

All numerical calculations have been performed using the resources of the Joint Supercomputer Center (JSCC) of RAS. The work was supported by the Program of the Presidium of RAS No. 21 "Foundations of fundamental investigations of nanotechnologies and nanomaterials".

## Notes and references

[a] *Institute of Problems of Chemical Physics of Russian Academy of Sciences, Chernogolovka, Moscow Region, 142432, Russia*

1   J. Baum, M. Munowitz, A. N. Garroway and A. Pines, *J. Chem. Phys.*, 1985, **83**, 2015-2025
2   C. E. Huges, *Progr. Nucl. Magn. Reson. Spectros.*, 2004, **45**, 301-313
3   H. C. Krojanski and D. Suter, *Phys. Rev. Lett.*, 2004, **93**, 090501
4   G. Cho, P. Cappellaro, D. G. Cory and C. Ramanathan, *Phys. Rev. B*, 2006, **74**, 224434
5   E. B. Fel'dman and S. Lacelle, *Chem. Phys. Lett.*, 1996, **253**, 27-31
6   E. B. Fel'dman and S. Lacelle, *J. Chem. Phys.*, 1997, **107**, 7067-7084
7   S. I. Doronin, I. I. Maksimov and E. B. Fel'dman, *J. Exp. Theor.Phys.*, 2000, **91**, 597-609
8   S. I. Doronin, E. B. Fel'dman, I. Ya. Guinzbourg and I. I. Maximov, *Chem. Phys. Lett.*, 2001, **341,** 144-152
9   W. X. Zhang, P. Cappellaro, N. Antler, B. Pepper, D. G. Cory, V. V. Dobrovitski, C. Ramanathan, L. Viola, *Phys.Rev. A.*, 2009, **80**, 052323
10  S. I. Doronin, A. V. Fedorova, E. B. Fel'dman, A. I. Zenchuk, *J. Chem. Phys.*, 2009, **131**, 104109
11  J. Baugh, A. Kleinhammes, D. Han, Q. Wang and Y.Wu, *Science*, 2001, **294**, 1505-1507
12  E. B. Fel'dman and M. G. Rudavets, *J. Exp Theor. Phys.*, 2004, **98**, 207-219
13  S. Lacelle, S.-J. Hwang and B. C. Gerstein, *J. Chem. Phys.*, 1993, **99**, 8407-8413
14  M. Tomaselli, S. Hediger, D. Suter and R. R, Ernst, *J. Chem. Phys.*, 1996, **105**, 10672-10681
15  U. Haeberlen, J. S. Waugh, *Phys. Rev.*, 1968, **175**, 453-467
16  M. Goldman, *Spin Temperature and Nuclear Magnetic Resonance in Solids*, Clarendon, Oxford, 1970
17  U. Haeberlen, *High Resolution NMR in Solids. Selevtive Averaging*, Academic Press, New Yourk, 1976.
18  E. B. Fel'dman and I. I. Maximov, *J. Magn. Reson.*, 2002, **157,** 106-113.
19  D. A. Lathrop, E. S. Handy, and K. K. Gleason, *J. Magn. Reson. Ser. A*, 1994, **111**, 161-168
20  L. D. Landau and E. M. Lifshitz, *Quantum Mechanics: Non-Relativistic Theory, Course of Theoretical Physics*, Vol. 3, Pergamon, New York, 1977
21  V. E. Zobov and A. A. Lundin, *Theor.Math.Phys.*, 2004, **141,** 1737-1749